# Challenges and complexities in application of LCA approaches in the case of ICT for a sustainable future

Reza FARRAHI MOGHADDAM,*§¶ Fereydoun FARRAHI MOGHADDAM,*§ Thomas DANDRES,†§
Yves LEMIEUX,‡§ Réjean SAMSON,†§ and Mohamed CHERIET*§

* Synchromedia Laboratory, École de technologie supérieure (ETS), University of Quebec (UduQ), Montreal, QC, Canada
† CIRAIG, École Polytechnique de Montréal (EPDM), University of Montreal (UdeM), Montreal, QC, Canada
‡ Ericsson Research Canada, Montreal, QC, Canada
§ CIRODD: P.O.Box 6079 Downtown, Montreal, QC, Canada, H3C 3A7
¶ Email contact of the corresponding author (Reza Farrahi Moghaddam): imriss@ieee.org

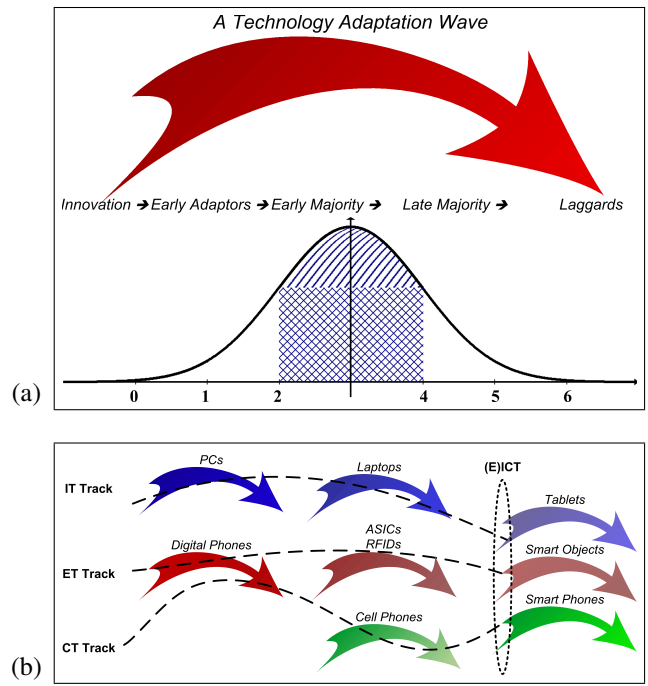

Fig. 1: A) a typical technology leap in the form of a technology adaption life cycle [2]. b) ICT as a meta-technology consists of various overlapping leaps along various tracks. Three tracks shown are Information (IT), Communications (CT), and Embedded (ET) [3]. Please see Supplementary Material S.A for more details.

*Abstract*—In this work, three of many ICT-specific challenges of LCA are discussed. First, the inconsistency versus uncertainty is reviewed with regard to the meta-technological nature of ICT. As an example, the semiconductor technologies are used to highlight the complexities especially with respect to energy and water consumption. The need for specific representations and metric to separately assess products and technologies is discussed. It is highlighted that applying product-oriented approaches would result in abandoning or disfavoring of new technologies that could otherwise help toward a better world. Second, several believed-untouchable hot spots are highlighted to emphasize on their importance and footprint. The list includes, but not limited to, i) User Computer-Interfaces (UCIs), especially screens and displays, ii) Network-Computer Interlaces (NCIs), such as electronic and optical ports, and iii) electricity power interfaces. In addition, considering cross-regional social and economic impacts, and also taking into account the marketing nature of the need for many ICT's product and services in both forms of hardware and software, the complexity of End of Life (EoL) stage of ICT products, technologies, and services is explored. Finally, the impact of smart management and intelligence, and in general software, in ICT solutions and products is highlighted. In particular, it is observed that, even using the same technology, the significance of software could be highly variable depending on the level of intelligence and awareness deployed. With examples from an interconnected network of data centers managed using Dynamic Voltage and Frequency Scaling (DVFS) technology and smart cooling systems, it is shown that the unadjusted assessments could be highly uncertain, and even inconsistent, in calculating the management component's significance on the ICT impacts.

## I. Introduction

In the long term vision toward a sustainable world in 2100, with its midterm checkpoint at 2050, dematerialization and its flag ship, i.e., virtualization, have been considered as key game changers [1]. ICT, as a meta-technology [S1][1] (Figure 1), seems to be a trivial tool to implement such concepts, and it has been projected that ICT would play an important role in virtualization of services even in a short term [4], [5]. In particular, it has been estimated that ICT will take a share of 14% of global electricity consumption (EC) by 2020 compared to its 4% and 4.7% shares in 2007 and 2012, respectively

[1] Because of the limited space, the citations marked with 'S' in the text are provided in the Supplementary References section of the Supplementary Material, which is accessible at http://arxiv.org/pdf/1403.2798.pdf#page=11.

[6], [7]. At the same time, ICT has been highly attractive to the service sector mainly because of its short mean time to deploy (MTTD) and also high return on investment [8], [S2], [S3]. It is foreseen that this would be exploited by current and expected service providers in near future leading to channeling most of the communication into ICT media, along with exponentially increase in both the number of interactions and also the number of connected devices and *things* to hundreds of billions [9]. All these scaling phenomena, short named as ICT enabling effect or ICT for a sustainable future, would raise a critical question on the possibility of unintentional harm from ICT to the world's sustainability in the case of

an even minor miscalculation in the decisions makings. This is a critical question because none of the aforementioned phenomena could have a significant impact except they are implemented and exercised at scales comparable to those of societies and economies, which would move them from being marginal into becoming mainstream. This confirms that decision makings for future, in both forms of policy making and also mass habit/behavior modification, are required to be based on validated and dependable assessments in order to prevent or at least minimize any unintentional harm.

If we look at the projected ICT's EC in 2020, i.e., 3,799 TWh, it shows an increase of more than 459% compared to the baseline of 2007, and it would be equal to 19.5% of the whole world's EC in 2012 [1], [7], [S4].[2] This is another concern about the ICT's stability considering the possible negative impacts of climate change on the security of energy sources across the world [10]. Along the same line, water consumption of ICT solutions and energy sources should be explicitly included in any sustainability analysis again because of the severe impacts of climate change on volatility of this scarce resource [11], [S5]–[S7].

The ICT sector usually is divided in three parts: Communication (Network), Devices,[3] and Data centers [7]. It has been observed that the EC of the devices part accounts for almost 57% of the total ICT's EC in the use phase in 2012. The associated share of the devices in the ICT's manufacturing EC has been more than 72% in the same year [12]. This shows a great opportunity to reduce the EC and footprint of ICT at high rates by cutting out the devices, replacing them with "thinner" devices, or smart management of their behavior (see section III-A for more discussions).

Network is also an important component of any ICT system [13], [14], especially the cellular networks that require a considerable physical deployment of infrastructure across wide areas. In an footprint analysis of cellular networks, it has been observed that, without proper utilization of the network, the manufacturing phase dominates the energy consumption and footprint [15]. This effect of underutilized infrastructure is common across all ICT systems, and we believe it is not only limited to the deployed infrastructure. Although increasing the utilization lifetime of equipment should be promoted, there is also a considerable volume of equipment that do not even get a chance to enter the market (see section III-D for more discussion).

Elasticity, which is one of the ICT potentials for sustainability, can be seen as the ability to dynamically scale the dedicated [ICT and support] resources[4] along time in response to the changes in the service requests. Elasticity has been known as an enabler of resource sharing, and usually is realized using resource virtualization. For example, in a study performed by Google, it has been observed that a high rate of 87% in EC saving can be achieved by virtualization and cloud computing used to eliminate dedicated servers [16].

[2] The ICT's EC in 2012 was 4.7% of the whole world's EC in that year [7]. [3] PC/OC/SP/TV (POST): Personal Computers/Office Computers/Smart Phones/TVs. [4] This includes at both manufacturing and operating phases.

However, it is worth mentioning that the rate of saving has been reduced to 34% when the "devices" are also considered in the calculations. This will be discussed more in section IV.

In this context, Life Cycle Assessment (LCA) has been suggested and considered for comprehensive calculation of environmental footprints and impacts, especially because of LCA's far-reaching nature (particularly along the life cycle of a product or a service to include the upstream and downstream processes) [17]. However, similar to the concerns about ICT's negative impacts on a sustainable future, there is a concern that decision making based on the not-fully-adapted-to-ICT assessments could result in delay in some of possible ICT's handprint toward a more sustainable world. To address these challenges, several international bodies, such as ITU, ETSI (EU), ISO, WRI, and WBCSD, among others, have been working to adapt and develop consistent standards of LCA for ICT [18]. Also, several testbeds, such as the Green Sustainable Telco Cloud (GSTC) project, have been deployed in order to validate the related models and assumptions. This paper is a small step along all these efforts to highlight some of critical elements and components that the authors believe require specific considerations in order to help in developing of dependable policies.

Various studies have estimated ICT systems' footprint using the LCA approaches [19]. A review of LCA for ICT is provided in [20] with an emphasis on possible missed impacts by considering whole life cycle and also social and behavioral aspects. Similar to other studies, they concluded that the manufacturing phase and the use phase have had the dominant share of the LCA impacts of many ICT systems. Another example is [21], where it was calculated that the embodied carbon footprint of a data center accounts for 51% of the data center's footprint. Moreover, support facilities of data centers and ICT solutions, especially cooling systems, represent a considerable portion of their footprint and EC [22] even if only the operating phase is considered [23].

Any improvement in the equipment lifetime would drastically enhance the performance of ICT solutions in terms of EC and footprint. However, it should not result in ignoring the hidden EC of home sensing and automation equipment, especially considering their large scale deployment in future in moves toward smart house and smart everything [24]. At larger scale of the whole Internet, it has been calculated that, among various devices, laptops have the highest ratio (=2.36) of embodied EC to wall-socket (use phase) EC [25]–[27], [S8]–[S11]. However, when the total [embodied plus wall-socket] EC of a typical laptop is compared to that of a typical desktop, an advantage of 40% can be estimated for the laptop. In other words, when the total life cycle EC is of great interest, its aggregation on the life stages may result in misleading conclusions because of the arbitrariness of reference. This suggests that the footprint of various stages should be presented in the form of a vector or other granular data representations, and the comparisons should be also made in the same way. Although this would require more reference data and use cases, the benefit would be much bigger because of availability of an absolute comparison.

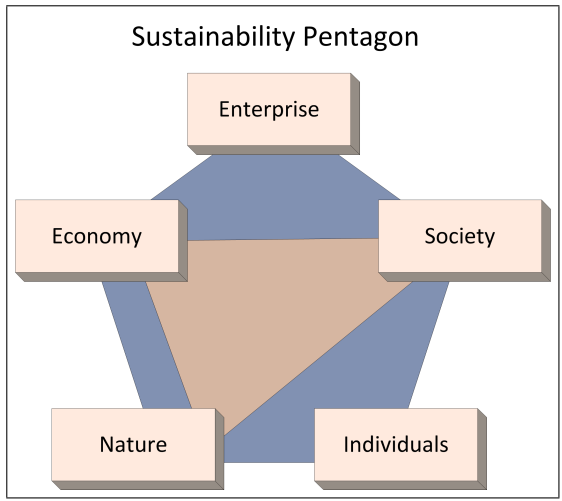

**Fig. 2:** The proposed revisited sustainability model.

LCA of electronics and ICT equipment has been studied for a long time, and resulted in standard databases [28]. However, it has been observed that consistency is still a big challenge [29]. Although more structured data and more samples are recommended as a way to improve the models, the proprietary nature of many manufacturing processes prevents access to accurate description of products. The consequential and rebound effects also add more to the complexity of analysis of ICT systems and their associated "green" or " sustainable" projects [30]. Especially, modeling the consumption trend and its evolution along time at multi-year scales could be a challenge and source of uncertainty [31].

It is worth noting that this work is not a review of LCA for ICT nor does it cover all elements of an LCA analysis. For reviews of the LCA work, please see [20], [32]. This paper tries to highlight a few ICT-specific challenges in assessment ranging from metric, to interfaces, and finally to software and intelligence. Because of lack of space, we skipped repeating those challenges that are common with other systems, such as rebound effect, system boundaries, open-loop recycling, among others. In addition, we introduce a more comprehensive sustainability model toward capturing various types of actors with conflicting interests and vulnerability.

Before discussing the LCA-specific challenges, we would like to mention a high-level challenge of the definition of *sustainability*. Although there have been various correct and valid definitions and models in the literature and practice, it seems that a new model is required in order to encompass all affected actors and entities considering the ongoing increase of the share of the sustainable actions and projects in the mainstream business-as-usual (BAU). A prospective model called the *sustaiblity Pentagon* is presented in Figure 2, and more details are provided in Appendix 1.[5] In a special case of application of such this new model, an LCA model is developed in Supplementary Material S.B.

## II. Challenge 1: Metric

Any analysis would face many specific challenges with ICT systems. In sections II to IV, we review three of these challenges, and we will discuss in section IV that the challenge related to software and intelligence is the most crucial and important one to be addressed. This is not a comprehensive list, and the purpose is to highlight those that would not appear in the analysis of non-ICT systems.

### A. Assessment of Product vs Assessment of Technology

As discussed before, ICT can be seen as a series of technological waves. To achieve ICT for sustainability, some decision makings are required to amplify the enabler waves. This would need a different approach to assessment compared to the assessment of products. For the latter, accurate estimation of its associated impacts is necessary. In other words, the potential of a technology used in a product cannot be claimed in order to monetize it by advertising or collecting credits, subsidies, or tax breaks.[6] In contrast, when estimating the benefits of a technology, relying on the impacts associated to its current realized products could be misleading. Especially, impacts associated to the place of manufacturing could compromise the potentials of an emerging technology even before its convergence to a stable state. An example is the Electrical Cars (EVs) case [33], [34], [S12], [S13], where the GHG emitted at mining the minerals required for the batteries and also the grid-associated emissions could outcast the emissions of gasoline cars. However, the conclusion and *decision* suggesting that EVs are not the way to go would be a miscalculation because it is not the EV technology itself that is responsible for the associated emissions but it is *placement* of its associated processes. Displacement or improvement can address these emissions without any change required in the core EV's technology. Another possible miscalculation would be to use the performance of a technology at the use phase to overcast the improper placement of the manufacturing/development phase, which prevents possible impact reducing actions especially at the regional levels. To add to the complexity associated to the spatial granularity aspect,[7] there are impacts that would greatly affect a local region compared to the rest of the world. For example, some types of air pollution in moderate levels would not transverse across regions mostly because of physical properties of the pollutant particles and geographical barriers.[8] Therefore there is a chance to improve regional policies even with negligible improvement in the total life cycle footprint.

### B. Uncertainty vs Inconsistency

It is well known that uncertainty is a big challenge for any assessment. However, in the case of ICT, there is a bigger challenge in the form of inconsistency, which would practically make assessments incomparable. This factor has attracted several researches toward standardization [36]. Inconsistency could partially arise from databases and models used to estimate the impacts. A famous example is the inconsistency between USA EIOLCA and EU PLCA life cycle inventory databases [36]. In addition, it was found that the LCAs of consumer electronics are not generally consistent [37]. For example, a difference of 341% in the estimations of the share of manufacturing in the total GHG footprint of devices were found between two analyses [37]. A similar conclusion was drawn in the case of desktop PC reflecting the high degree of variability in terms of their performance and power consumption [38], [39]. Also, confidentiality of manufacturing data in the semiconductor industry has added to complexity because even the degree of the inconsistency between two products or components could not be determined [40]. Another source of inconsistency would be inclusion or omission of Total Cost of Ownership (TCO), which includes maintenance, training, and management. It has been observed that the TCO can contribute in an additional 86% EC in a use case associated with the Internet [41]. Also, temporal aspect of ICT systems, especially in terms of long-term trends, adds additional inconsistency, and requires a better understanding [42].

---

[5] The details of the new sustainability model will be provided in another work.

[6] Other ways to monetize sustainability-driven projects could be targeting *individuals* (behavior) and *society* (subsidizes) in addition to regular benefits from reduction in resource consumption and long-term liabilities. [7] Although some of impacts, such as GHG emissions, are eventually shared by all regions. [8] Although it has been observed that they could initiate different types of climate change in other regions [35].

The main source of inconsistency could be seen in high degree of variability, even within a single technologies. Moreover, as will be discuss in section III, a large portion of impacts and therefore inconsistencies would reside on the *thin* interfaces and borders within an ICT system. It is usually argued that to some extend the impacts of a product can be inferred from its group, functionality, and type [43], [44]. Although this would suggest to consider parameterization and development of parametric models in order to cut assessment costs and improve the insights [45], caution should be exercised because impacts sourced from interface and border regions of the system could easily escape the analysis, or could disturb its certainty and consistency, especially if linear-relation models between mass and embodied footprint are practiced [46]. These escapes might be exploited as loopholes in prospected regulations or policies, similar to those misuses reported in the case of the Ozone Depleting Substances (ODS) regulations [47].

Uncertainty is a challenge however at the same time it is a core part of any analysis. In contrast, inconsistency is rooted in incapability of the models to capture the complete picture. The disparities in the models that propagate and get amplified along the analysis process cannot be addressed by uncertainty analysis because they are independent from the amount of data collected. An example would be comparing two technologies while their place of manufacturing is different. This comparison is inconsistent even if the same unit of service/product delivered is considered in both cases. The argument that some technologies are place-dependent in their manufacturing/development is gradually fading out, mostly because of the *knowledge without borders* phenomena powered by mobility of scientists across regions [48].

## C. Use case: Semiconductor technologies

As a use case, the EC and water footprint of semiconductor technologies is discussed here. In a high-level classification, these technologies can be classified according to their average half-pitch[9] [49]. Considering 7 technologies from 350-nm to 45-nm, it was observed that upstream energy consumption is decreasing with new technologies, while that of use phase increase with new and smaller technologies mainly because of their higher power demand [49]. In particular, the energy consumption per die increases with a power-law $n$-value of 1.13 from 151 MJ at 350-nm to 1,593 MJ at 45-nm in the use phase.[10] The main difference between that study and many other studies in the same direction is the way the lifetime of equipment is handled. In contrast to traditional notion of a-few years (for example, 5 years) lifetime assumption, [49] assumes specific utilization hours for the lifetime. Although these two forms should be in general equivalent, the latter allows for utilization "beyond" the nominal lifetime. We will come back to this aspect in section III-D to mention concerns and possibilities with respect to the equipment lifetime "extension."

At the same time, it has been observed that the EC in manufacturing phase of processors is almost constant at an average of 3kWh per unit [51]. However, considering various support and secondary functions required, such as heating, cooling and ultra-pure water, there is also a large room for impact reduction and optimization in the manufacturing phase [52]. In terms of life cycle water use, a decreasing power-law relation with an $n$-value of 0.69 can be calculated, where 10,206 liter of water per die is associated with the 45-nm use phase representing 97% of its total water use,[11] and can be traced back to electricity generation [49].

This way of analysis, which separates the place of manufacturing from the technology itself, helps in more direct decisions. In particular, the dominance of the use phase EC and water use suggests to consider i) smart and dynamic approaches to power management (see section IV-D for an example), ii) proper selection of technology based on the application [50], [53], and iii) increase in the utilization lifetime. Although the latter seems to be contradictory with the "immediate" goal of impact reduction, considering higher power demand associated to old equipment, adverse effects of any delay, especially across developing regions, would have much more costly unsustainabilizing consequences (see section III-D). Increasing the lifetime of ICT equipment and services in anyway, even in the form of displacement across regions, would help to increase their access and therefore global development and stability.

To put this is a formulation, we propose that the footprint of an ICT system is expressed as an 1-dimensional array $\mathbf{F} =$

---

[9] i.e., half the distance between identical features. [10] Calculated using a 6,000 hours lifetime assumption, a 70% power supply efficiency, and considering 14W and 146W power demand for 350-nm and 45-nm technologies, respectively [49]. The use phase EC of 45-nm represents 94% of its total energy consumption. This is consistent with other studies in which, for example, ultra-low-power MCU and power-demanding 32-nm CPU technologies consume 4.7mW and 42W in active mode, respectively [50].
[11] When modified manufacturing processes are used [49].

$\{F_i\}_i$, where $F_i$ ($F_i \geq 0$) is its associated *downward* footprint on nature(s) at $i^{th}$ stage of the *utilization lifetime*. The $F_i$ can be seen as a collection of various *attributes* describing the footprint. Referring to Figure S-1, an additional dimension could be considered to take into account *upward* interactions with societies: $\mathbf{F} = \{F_{i,j}\}_{i,j}$, where $j \in \{\text{d}, \text{u}\}$ counts on downward footprint and upward handprint.[12] The main part of formalism is the definition of a proper *distance* function that allows highlighting minor positive changes even in presence of major stationary offsets. Although we will address this aspect in future work, a generalization of the Canberra distance [55] seems to be promising:

$$\text{d}\left(\mathbf{F}^1, \mathbf{F}^2\right) = \left\{\sum_i \left|\frac{F_i^1 - F_i^2}{F_i^1 + F_i^2}\right|_\omega\right\}_\omega, \quad (1)$$

where $|\cdot|_\omega$ is the distance associated with the attribute $\omega$.

## III. CHALLENGE 2: INTERFACES AND END-OF-LIFE

As mentioned before, it seems that a big portion of footprint associated to the ICT solutions is sourced from the "interface" regions between various physics constituents involved in these systems. In particular, we will discuss three types of interfaces in this work: i) Human-Computer Interface (HCI), ii) Network-Computer Interface (NCI), and iii) Power-Computer Interface (PCI). In short, we call them HNP-CI. Other interfaces could be added to this list in the future. The reason that interfaces could have a high potential to degrade the performance and increasing the footprint can be traced to their *multi-physics* nature. Understanding of multi-physics phenomena is still a challenge, and the associated knowledge is mostly confidential and is held by a few actors. This "low mass" has possibly prevented acceleration and has also lowered motivation in the associated R&D. The other important factor would be "lack" of communication means between the two physics of an interface. Lack of knowing the state of the other side could force one side of an interface to be always in an "active" mode and ready to provide even in the *absence* of the other side. This shows a great opportunity to reduce the EC by increasing intelligence and also improving communication across interfaces.

### A. Human-Computer Interface (HCI)

Display screens of PC/OC/SP/TVs (POST) have been recognized as a major contributor to the EC and also one of main factor in shortening BAU-lifetime of devices. The HCIs have been usually considered as one way interfaces. With the move toward interactive and two-way HCIs, there is a high potential to add intelligence to these energy hungry components of the ICT solutions in order to dynamically adjust their operation especially according to temporal changes in their reachedness by the second side, i.e., human. In addition, deeper penetration of computer side in the human premise seems to enable drastic reduction in resource use, EC and its associated footprint

---

[12] Although handprint is usually considered as positive impact [54], we here consider it to represent any positive or negative upward interaction.

[56]. For example, smart glasses and contact lenses can be mentioned, which can be considered in the category of the Interface-to-the-Eye (IttE) penetration.

### B. Network-Computer Interface (NCI)

The same argument holds for the NCIs. These *ports* have complex multi-physics at high speeds. Despite this fact, collaborative operation of multiple ports within a smart management could allow turning many ports off. Roughly, a power reduction from 0.5W (disabling a port) to 30W (turning off a whole switch) per port could be tapped into [57].

### C. Power-Computer Interface (PCI)

PCIs are the most difficult interfaces because of high volume of energy passing through them. In contrast to other interfaces, the main strategy to reduce the EC associated to the PCIs is eliminating them. One of promising approaches has been Direct Current (DC)-powered ICT systems. The DC nature of batteries and also many ICT equipment allows completely-DC architectures in which inverters and converters are only present at the border with the external electricity sources, such as electricity grids. At the level of the electricity grid, Power-Power Interfaces (PPIs) are important components especially at the interconnects/intraconnects with renewable wind and solar power plants. Therefore there is a great potential to improve the PPIs' performance by considering the electricity grid itself as a giant ICT system [58], [S14], [S15].

### D. End of Life Challenges

End of Life (EoL) has its own challenges in LCA for ICT, especially considering unavailability of information on the actual approach used to recycle or dispose e-wastes. Therefore, special attentions are required to handle potential secondary effects. As mentioned in section II-C, traditional BAU to decommission equipment after a few years, which probably roots in lax and cheap [material and labor] resource availability, have drastically affected the results of many assessments. Although adaptation of this BAU in the assessments seems a correct modeling action, we suggest that the baseline should be moved to the "potential" lifetime instead of the BAU lifetime. In other words, if an equipment, with a potential lifetime of 10 years, is terminated at the third year, the additional footprint associated to the missing 7 years should be added: This would include 233% increase in the upstream and downstream footprint. Also, as mentioned before, increasing the intelligence and also shifting the *tempting* aspects of the ICT solutions to software components (such as Apps) could help to increase the potential lifetime of equipment.

*1) Obsolescence and E-waste: Planned or unavoidable:* It seems that short BAU lifetime of equipment and devices could be possibly rooted in the interest of manufactures and service providers, and therefore the associated footprint should be also reflected in their assessments, and eventually reduced by proper policies. Before that time, various approaches could be used to practically extend the lifetime. In addition to recycling, reuse seems to be a proper temporary approach especially in multi-region scenarios [59], [60], [S16], [S17].[13] In addition, refurnish and repair actions are also highly recommended in particular in cross-sector scenarios, for example from Telco to education. Although these approaches would be temporary and would work only for current equipment in use,[14] proper consideration in "look and feel" of future equipment would help them stay longer in operation even in saturated markets [S17], [S18].

## IV. CHALLENGE 3: SOFTWARE AND INTELLIGENCE

In contrast to many other industry sectors, software plays a critical role in the ICT solutions. In short, an ICT infrastructure is inoperable if the software component is dropped, which means software should be practically considered as a Part of Infrastructure (PoI). This has been correctly observed by many LCA standards and methodologies for ICT [18]. However, there are enormous number of challenges beyond this level needed to be considered mostly because of variability, interdependency, and legacy-lingering nature of software components. In this section, a few of these challenges, some of which can be seen as "side effects" of green actions, are discussed as examples. Later on, the big potential of software component in increasing the efficiency, reducing the footprint, and in general improving the performance will be discussed and a use case corresponding to the DVFS technology will be presented in section IV-D.

### A. Adverse effect 1: The example of virtualization

Although ICT seems to be a powerful enabler toward dematerializing, preventing resource depletion and also avoiding pollution, its technologies that actually deliver such possibilities could impose negative, secondary impacts.[15] For example, virtualization class of technologies has been promoted as an enabler of resource sharing and resource independency within the ICT industry. This is mainly thanks to virtualization's inherent nature of abstraction [61]. There are many approaches to virtualization. Although almost all these approaches are continuously evolving thanks to their growing developers' communities, it is worth to mention some of their current side effects.[16] An example is the overhead associated to hypervisors and virtual machine managers (VMMs). With recent advances, the performance of these fundamental components of virtualization has drastically improved with respect to CPU and memory.[17] However, their main bottleneck is still the I/O and networking performance [63], [64], [S19]. It has been observed that hypervisors introduce a considerable amount of network overhead. This should be combined with the overhead related to service discovery and also protocol translations, which are essential for abstraction [65]. Even if we assume that the direct overhead of virtualization components is hypothetically negligible, there are other hidden secondary overheads that should be accounted for. For example, a common aspect of virtualized system is their tendency toward more abstract layers of network, such as network layer (layer 3) with the IP protocol. It has been reported that the software processing accounts for almost 72% latency (averaged on transmit and receive stages) in the TCP/IP communications. This would suggest that adaptation of less compute-intensive architectures or protocols, such as InfiniBand (IB) or 9P, would help to increase the performance and also to reduce the associated secondary footprint associated to extra software processing [66], [67]. Also, compromise between bandwidth and latency could highly affect the amount of secondary software processing involved, and in turn the amount of associated impact [68]. The latency factor, which is critical to many applications such as those of Telco, could impose a constraint on the possibility of reducing these secondary impacts especially considering the small size of messages that are required to be exchanged. This again supports our argument on the necessity to represent the impacts in the form of granular vectors or arrays in order to preserve the accountability of the secondary impacts, which otherwise would be simply trimmed when combined with primary impacts. Although the total footprint of the system would not change, more granular footprint representations allow decision making at the same low levels of granularity, which could be otherwise impossible to be achieved because of independency of benefits and costs of various actors involved. A specific example, is the authentication messaging in IP Multimedia Subsystem (IMS) systems considered in the GSTC project. Before establishing a media link between the two parties of a session probably using Voice over Internet Protocol (VoIP) technologies, small but many Session Initiation Protocol (SIP) messages are exchanged between the IMS servers and the callers. In an aggregated impact picture, the footprint of these messages, which are in order of KBs of data exchange, is simply negligible compared to media exchanged in volumes of several MBs. However, the proposed granular impact representation preserves these small impacts in the assessment, which in turn could be leveraged to initiate decisions, for example, at the level of the IMS system developers who would be otherwise unchallenged in an aggregated approach.

### B. Adverse effect 2: Management of the VM images

Another aspect of virtualization, as an enabler of resource share, is the challenge of management of the Virtual Machine (VM) image files. Virtualization allows abstraction and separation of most of the software part in the form of VMs, which also allows re-startable "snapshots" of the system state in the form of VM image files. However, the number of these images could drastically increase even only for one system.

---

[13] Cautions should be exercised because of health issues related to old equipment in the [unregulated] destination regions, and also increased consumption (rebound effect) in the source regions. [14] Depending on when the destination regions or sectors are saturated. [15] Although these negative impacts would be negligible when compared to the positive benefits of the associated dematerialization, they would be considerable when they are used in the decision makings to choose between two competitive ICT technologies that deliver the same level of dematerialization. Decision makings are very important because ICT technologies usually induce a high-degree of legacy that would require a total reformation if a decision is required to be reverted.
[16] The authors think that virtualization technologies would be a mainstream toward a sustainable future, and the mention of side effects here is to initiate seeking modifications and solutions in order to address and minimize them.
[17] Only 10% overhead is observed for compute-intensive jobs [62].

Therefore, storage and management of these images could impose a considerable infrastructure requirement and therefore footprint [69], [S20]. On top of that, there is a secondary effect of "startup" overheads. In particular, it has been shown that the size of VM image files has a big impact on the launch time [70], and therefore on the associated software processing and also network communications. Primary (storage and management) and secondary (startup overheads) of VM images should be critically considered. Promoted by the granular impact assessment, a promising decision would be reducing the size of VM images and also the consolidation of VM images based on their similarities [71].

*C. Intelligence*

Although the Management/Intelligence (M/I) aspect of an ICT solution is realized mostly in the software form, its impact on the overall performance and footprint could be so high that special attention is required. This is of great interest because "minimal"-impact ICT systems are considered as candidates toward smart everything, resource preservation, and footprint reduction in all other sectors [72]. An example of the substantial impact of M/I on the footprint of a distributed network of data centers will be provided below.

*D. Experimental Results: The use case of Dynamic Voltage and Frequency Scaling (DVFS)*

As mentioned in section IV-C, intelligence can drastically affect the performance of an ICT solution, and therefore it requires a special attention in assessment and reporting. To put this in concrete words, a use case based on the DVFS technology is presented. DVFS has been an attractive feature in the smart management of ICT systems, especially compute nodes [23], [50]. By definition, DVFS stands for the ability of the hardware, mostly CPU processors, to allow adjusting their clock frequency by software in real-time [73], [74], [S21]. This is of great interest because there is a power law ($n$-value of about 3) relation between the frequency and power requirement of a processor, and therefore there is a high opportunity to reduce the EC by adjusting the frequency [23]. We will show with the results of this use case that "selection" of the M/I used for this purpose would mainly determine the actual saving in the EC. Depending on the nature of application, the problem would result in an NP-hard combinatorial problem usually only solvable using greedy, soft-greedy, or heuristic approaches [23].

A simple DVFS model of a CPU processor's power can be expressed as follows:

$$P_{CPU,f}(t,\mathbf{l}) = \left(\frac{1}{n_{\text{cores}}}\right) \times \sum_{i=1}^{n_{\text{cores}}} \left(P_0' + \left\{P_1'\frac{f_i(t)}{f_{\max}} + P_2'\frac{f_i^2(t)}{f_{\max}^2}^2 + P_3'\frac{f_i^3(t)}{f_{\max}^3}\right\}l_i\right). \quad (2)$$

After excluding first and second degree terms, we have:

$$P_{CPU,f}(t,\mathbf{l}) = \left(\frac{1}{n_{\text{cores}}}\right) \sum_{i=1}^{n_{\text{cores}}} \left(P_0 + P_3 \frac{f_i^3(t)}{f_{\max}^3} l_i(t)\right), \quad (3)$$

where $l_i$ is the load ratio on the $i^{th}$ core of the CPU defined as the ratio of the instantaneous rate of MIPS[18] executed on that core divided by its nominal MIPS at frequency $f_i(t)$, and $\mathbf{l} = \{l_i | i = 1, \cdots, n_{\text{cores}}\}$. $f_{\max}$ stands for maximum achievable frequency of the processor clock. The temporal load ratio of the whole CPU processor is defined as: $l(t) = \sum_i l_i(t)/n_{\text{cores}}$. For example, we can obtain $P_0 = 142.2W$ and $P_3 = 107.8W$ for the case of a 250W-3GHz server [23], [75].

In this use case, we assume an HPC distributed network of data centers at various geographical places across the globe equipped with DVFS-enabled CPUs. Each data center has a smart cooling system, which is fully parameterized and is controllable via a control vector of utilization rate of its CRAC,[19] chiller plant, and cooling tower [23]. This allows the system to benefit from low-temperature hours at various places distributed across several time zones, and also from dynamic adjustment of the cooling power depending on the actual load. It is observed that this permits lowering the Power Usage Effectiveness (PUE) of the system to 1.139 down from the original baseline of 1.454 whereby the cooling systems were not optimized. A cooling system can greatly influence the EC and footprint of a data center [23], [76], [77], [S22].

To show the impact of M/I on the EC and footprint, several DVFS-aware M/I algorithms were considered on the same infrastructure and with the same load trace. Table I shows the performance of these algorithms against each other and the proposed Carbon-Profit-Aware scheduler algorithm (for duration of a week). It is worth mentioning that the proposed algorithm is set to optimize the profit of the system in contrast to other algorithms that target only emissions or EC. This is important because a non-profitable system does not fulfill sustainability requirements even if it has zero footprint. As mentioned before, despite having identical systems that all use DVFS, the EC has a large variations mainly because of the selection of the M/I algorithm: The EC of the use phase has a 71% increase in the case of the ENER algorithm compared to that of the CARB algorithm. Therefore, assessments that ignore the M/I influence of the ICT solutions would face a considerable amount of uncertainty that is independent from the material (weight) used. Another interesting point is when we compare the proposed CPAS algorithm with the ENER algorithm. Although their EC is almost the same, the profit of the CPAS case is 28% higher, which shows profitability does not always mean higher consumption.

## V. CONCLUSIONS AND FUTURE PROSPECTS

ICT is a preferred candidate for transition to a sustainable future. However, there are various challenges and aspects that, if not properly addressed, could eventually cancel out the benefits of ICT. In this work, three ICT-specific challenges among many others have been highlighted and discussed. First, the necessity of new representations and metric to evaluate and compare various competitive ICT technologies has been highlighted, and a prospective metric has been suggested. Second, lifetime extension and also importance of interfaces in the ICT systems have been discussed. Interfaces could have a considerable amount of hidden footprint. Third, the drastic impacts of software and intelligence on the performance, consumption and footprint of ICT solutions have been presented. A use case of the DVFS technology, as an example, has been provided to illustrate that high variations (up to 70%) can be observed in the footprint assessments just by changing the management software and intelligence algorithm while leaving the rest of the system (including the hardware and the rest of software) untouched. In addition, a revisited sustainability model has been mentioned that covers all actors involved in large scale sustainability activities with paying special attention to multi-region and multi-interaction behaviors. Every aspects mentioned in this paper requires more comprehensive analyses in the future.

---

[18] Million Instructions Per Second.   [19] Computer Room AC (CRAC).

| *Algorithm* / Metric | PERF | ENER | CARB | PROF | CPAS (Proposed) |
|---|---|---|---|---|---|
| Use EC (kWh) | 23,473 | 20,877 | **12,197** | 23,031 | 19,727 |
| GHG (kgCO$_2$e) | 14,140.98 | 12,401 | **7,276.2** | 14,010 | 12,270 |
| Profit ($) | 1,151.7 | *1,438.2* | 1,010.8 | 1,373.0 | ***1,837.8*** |
| Frequency (GHz) | 3 | 2.79 | 1.84 | 3 | 2.68 |

**TABLE I:** A comparison of the impact of M/I. Acronyms: Performance-based scheduler (PERF) [78], Energy-based scheduler (ENER) [79], Carbon-based scheduler (CARB) [80], Profit-based scheduler (PROF) [81], and proposed Carbon-Profit-Aware scheduler (CPAS) [23].

## Acknowledgment

The authors thank the NSERC of Canada for their financial support under Grant CRDPJ 424371-11 and also under the Canada Research Chair in Sustainable Smart Eco-Cloud.

APPENDIX 1. NEED FOR A NEW SUSTAINABILITY MODEL

When actions and interactions, driven from a change, scale up and become considerable versus the baseline volume of actions involved in a system or an ecosystem, the associated model is also required to be accordingly adjusted. The case of nature-aware or nature-friendly[20] changes is our focus in this work. When a proposed "green" change or action is small scale, nature-friendliness is a sufficient approximation because the rest of the system (or the world) could be considered as a reservoir unaffected by such actions, and therefore a micro-canonical analysis is correct. However, in recent years, with scaling up of the *green* actions, especially in the domain of energy sources and electricity generation, the aforementioned assumption is no longer valid, and this has naturally resulted in considering a more complex sustainability model that also includes the society and economy, and addresses the eco-socio-environmental footprint [82], [S23], [S24]. This triangular model, shown also in Figure 2, has been used to assess changes or actions in terms of their sustainability by projecting their impacts on these three complementary aspects.

We suggest to introduce a new sustainability model to *stimulate* actions, in addition to evaluating their impact (shown as a pentagon in Figure 2). This prospective model requires to encompass all players and actors to understand them and then to involve them in the sustainability *game*.[21] The actors are classified based on their interest and also their actions. We suggest 5 categories of actors in this model. The two additions cover *individuals* and *enterprises*. Although these actors are practically also residing within the society actors, their actions and especially their interest could drastically diverge from those of the society(ies), and therefore separated categories are considered. To be more accurate, at the actions level, the individual category represents the actions of every individual, while the enterprise one represents collective actions of every group of individuals, and finally the society category represents collective actions of all individuals bounded to a specific region or characteristic. Similarly, the interests of each category is defined. This Sustainability Pentagon is a framework to represent interactions (including impacts) among actors from each of these five categories. The actors dynamically take various roles of subject, object, and partner in this model depending on the nature of actions involved among them.

Also, it is worth noting that, in contrast with earlier models, the 5 categories are *multivalent*. Although this is trivial for the individuals (people) and enterprises (businesses), it explicitly should be considered and respected for the other categories especially because of the global interactions involved and also because of fundamental, dynamic differences among societies implicated [84], [85]. Even, the nature and economy categories could be split across the regions. As an specific example, the *atmosphere* could be mentioned. Although the atmosphere is usually considered as shared among all regions, and therefore it is imagined as a *single* natural resource and object with respect all GHG emissions, it would be split into many sub-objects under the nature category when spatially-contained impacts, such as smog and air pollution, are concerned.

---

[20] Also, the term *environment* has been causally used in the place of *nature* in the literature. However, with introduction of new and more complex "environments" that also encompass human, caution should be exercised to avoid confusion. Digital Environment and Electromagnetic Environment are a few examples.

[21] Modeling multi-actor systems and behavior management using game theory have recently attracted a considerable interest [83], [S25].

SUPPLEMENTARY MATERIAL

*A. A Discussion on the ICT as a Meta-Technology*

Here we discuss in more details the technological waves of ICT that have shown in an informal way in Figure 1(b). Although technological waves usually have a definite period of time, they could highly overlap each other. In addition, various technological waves could be clustered along specific paths or tracks. Therefore, we used three tracks in the figure to present several waves in the same picture.

In particular, three tracks of Information Technology (IT), Communications Technology (CT), and Embedded Technology (ET) have been considered. For the IT and CT tracks some of classical examples are presented in the figure, while for the ET track we used two specific examples of Application Specific Integrated Circuit (ASIC) [S26], [S27] and Radio-Frequency Identification (RFID) [S28]. Although these tracks have had been considered to be completely diverse in the past, it seems they have been converging along each other toward a single mega track. This convergence, which has been marked as ICT,[1] could provide solutions and means toward a unified transition to the era of smart objects and Internet of Things. The detailed discussion on the associated capabilities and impacts is beyond the scope of this work.

*B. Sustainability Pentagon and the Associated LCA Model*

One way to approximate the revisited sustainability model of Figure 2 is to *flatten* it along the *utilization lifetime* of a product or service. The flattened approximation, which resembles a life cycle analysis, allows a systematic cutout of actors that have little or negligible involvement in the target product/service. An example of the flatten model is illustrated in Figure S-1. The individuals (people) and enterprises (businesses) involved are all placed in the same slab, while the actors from the society and nature categories are placed in their own specific slabs extended in parallel along the life span.

SUPPLEMENTARY REFERENCES

[S1] T. Rayna, L. Striukova, and S. Landau, "Crossing the chasm or being crossed out: The case of digital audio players," Int. J. of Actor-Network Theory and Technological Innovation, vol. 1, no. 3, pp. 36–54, 2009.

[S2] S.-Y. T. Lee, R. Gholami, and T. Y. Tong, "Time series analysis in the assessment of ICT impact at the aggregate level — lessons and implications for the new economy," Information & Management, vol. 42, no. 7, pp. 1009–1022, Oct. 2005.

[S3] A. Emrouznejad, E. Cabanda, and R. Gholami, "An alternative measure of the ICT-opportunity index," Information & Management, vol. 47, no. 4, pp. 246–254, May 2010.

[S4] International Energy Agency, "World energy outlook 2020 and global electricity investment challenges," International Energy Agency, Paris, Île-de-France, France, Tech. Rep., 2002.

[S5] M. Berger and M. Finkbeiner, "Water footprinting: How to address water use in life cycle assessment?" Sustainability, vol. 2, no. 4, pp. 919–944, 2010.

[S6] B. G. Ridoutt and S. Pfister, "A revised approach to water footprinting to make transparent the impacts of consumption and production on global freshwater scarcity," Global Environmental Change, vol. 20, no. 1, pp. 113–120, Feb. 2010.

[S7] R. Hernandez, S. Easter, M. Murphy-Mariscal, F. Maestre, M. Tavassoli, E. Allen, C. Barrows, J. Belnap, R. Ochoa-Hueso, S. Ravi, and M. Allen, "Environmental impacts of utility-scale solar energy," Renewable and Sustainable Energy Reviews, vol. 29, no. 0, pp. 766–779, Jan. 2014.

[S8] V. C. Coroama, L. M. Hilty, E. Heiri, and F. M. Horn, "The direct energy demand of Internet data flows," Journal of Industrial Ecology, vol. 17, no. 5, pp. 680–688, 2013.

[S9] V. C. Coroama and L. M. Hilty, "Assessing Internet energy intensity: A review of methods and results," Environmental Impact Assessment Review, vol. 45, pp. 63–68, 2014.

[S10] K. Hinton, J. Baliga, M. Feng, R. Ayre, and R. Tucker, "Power consumption and energy efficiency in the Internet," IEEE Network, vol. 25, no. 2, pp. 6–12, 2011.

[S11] A. Moberg, M. Johansson, G. Finnveden, and A. Jonsson, "Printed and tablet e-paper newspaper from an environmental perspective — a screening life cycle assessment," Environmental Impact Assessment Review, vol. 30, no. 3, pp. 177–191, 2010.

[S12] D. S.-K. Ting, "Book review. green illusions: the dirty secrets of clean energy and the future of environmentalism," International Journal of Environmental Studies, vol. 70, no. 5, pp. 832–835, May 2013.

[S13] P. Newell and D. Mulvaney, "The political economy of the 'just transition'," The Geographical Journal, vol. 179, no. 2, pp. 132–140, 2013.

[S14] S. D. Ramchurn, P. Vytelingum, A. Rogers, and N. R. Jennings, "Putting the 'smarts' into the smart grid: A grand challenge for artificial intelligence," Commun. ACM, vol. 55, no. 4, pp. 86–97, 2012.

[S15] S. Rohjans, C. Danekas, and M. Uslar, "Requirements for Smart Grid ICT-architectures," in ISGT Europe'12, Berlin, Berlin, Germany, October 14-17 2012, pp. 1–8.

[S16] A. Martinuzzi, R. Kudlak, C. Faber, and A. Wiman, "CSR activities and impacts of the ICT sector," Vienna University of Economics and Business, Vienna, Wien, Austria, RIMAS Working Paper 5/2011, 2011.

[S17] E. Williams, R. Kahhat, B. Allenby, E. Kavazanjian, J. Kim, and M. Xu, "Environmental, social, and economic implications of global reuse and recycling of personal computers," Environ. Sci. Technol., vol. 42, no. 17, pp. 6446–6454, Aug. 2008.

[S18] P. Dourish, "HCI and environmental sustainability: the politics of design and the design of politics," in DIS'10, Aarhus, Midtjylland, Denmark, August 16-20 2010, pp. 1–10.

[S19] E. Shoop, R. Brown, E. Biggers, M. Kane, D. Lin, and M. Warner, "Virtual clusters for parallel and distributed education," in SIGCSE'12, Raleigh, NC, USA, February 29 - March 3 2012, pp. 517–522.

[S20] B. Nicolae, J. Bresnahan, K. Keahey, and G. Antoniu, "Going back and forth: efficient multideployment and multisnapshotting on clouds," in HPDC'11, San Jose, CA, USA, June 8-11 2011, pp. 147–158.

[S21] T. Furuyama, "Deep sub-100nm design challenges," in ASSCC'06, 2006, pp. 7–10.

[S22] D. J. Lettieri, "Expeditious data center sustainability, flow, and temperature modeling: Life-cycle exergy consumption combined with a potential flow based, Rankine vortex superposed, predictive method," Ph.D. dissertation, University of California, Berkeley, Berkeley, CA, USA, Spring 2012.

[S23] D. Griggs, M. Stafford-Smith, O. Gaffney, J. Rockstrom, M. C. Ohman, P. Shyamsundar, W. Steffen, G. Glaser, N. Kanie, and I. Noble, "Policy: Sustainable development goals for people and planet," Nature, vol. 495, no. 7441, pp. 305–307, Mar. 2013.

[S24] B. Hopwood, M. Mellor, and G. O'Brien, "Sustainable development: mapping different approaches," Sust. Dev., vol. 13, no. 1, pp. 38–52, 2005.

[S25] A. G. Sanfey, "Social decision-making: insights from game theory and neuroscience," Science, vol. 318, pp. 598–602, 2007.

[S26] S. Tewksbury, "Application-specific integrated circuits (ASICS)," in The Electrical Engineering Handbook, R. Doff, Ed. CRC Press, 1998.

[S27] N. Farrington, G. Porter, S. Radhakrishnan, H. H. Bazzaz, V. Subramanya, Y. Fainman, G. Papen, and A. Vahdat, "Helios: a hybrid electrical/optical switch architecture for modular data centers," SIGCOMM Comput. Commun. Rev., vol. 40, no. 4, pp. 339–350, 2010.

[S28] C. Roberts, "Radio frequency identification (RFID)," Computers & Security, vol. 25, no. 1, pp. 18–26, 2006.

---

[1] It could be also extended to explicitly include embedded-enabled devices: (E)ICT.

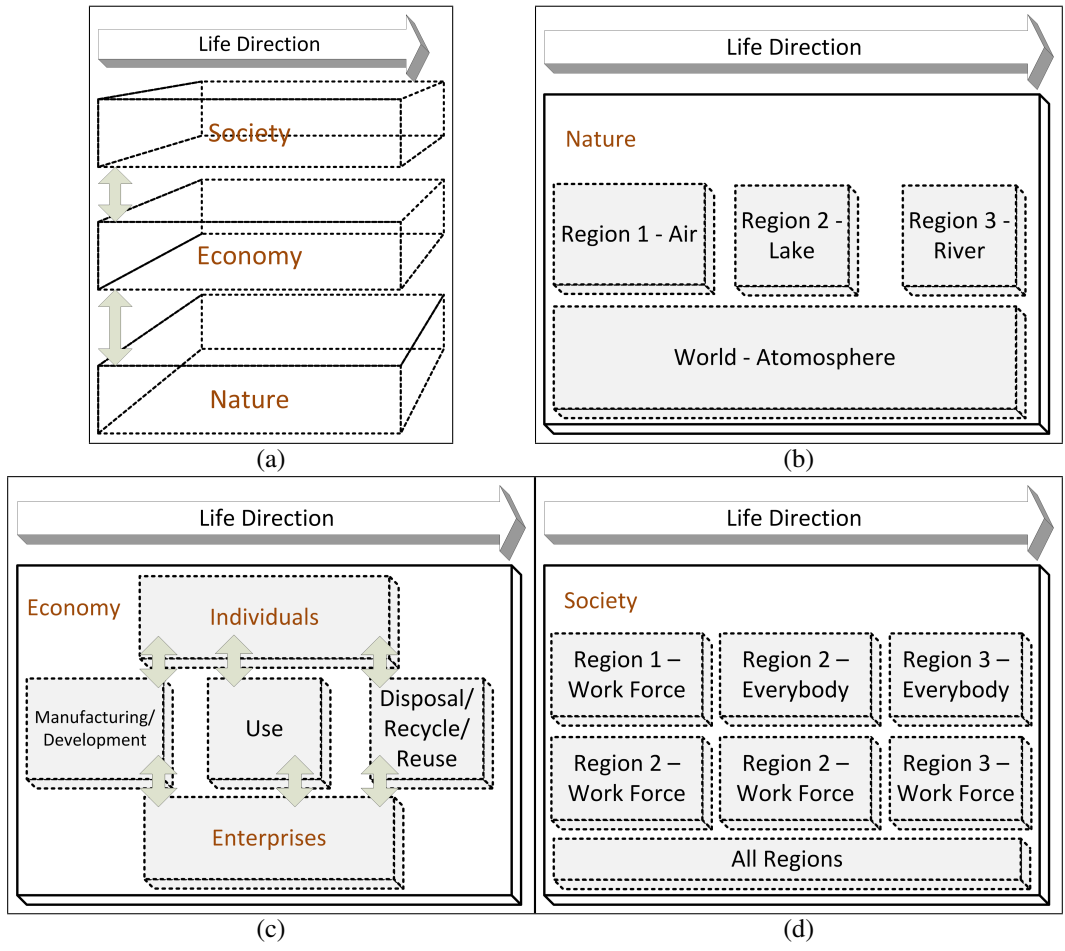

**Fig. S-1:** a) The flattened approximation of the sustainability model. b)-d) The exemplary detailed three slabs of (a).